\documentclass[conference]{IEEEtran}

\usepackage{graphicx} % Allow \includegraphics
\usepackage{tabularx} % The more x, the better
\usepackage{float} % Allow \restylefloat
\restylefloat{table} % Allow tables to use the H placement
\usepackage{changepage} % Allow adjustwidth environment
\usepackage{amsmath} % Allow \dfrac
\usepackage{wrapfig} % Allow wrapping figures
\usepackage{url} % \url
\usepackage{listings} % source code highlighting
\usepackage[frozencache]{minted} % Source code highlighting, better?
\usepackage{hyperref}

\newcommand{\enquote}[1]{``#1''} %enquote stuff nicely
\newcommand{\multicell}[2][t]{\begin{tabular}[#1]{@{}l@{}}#2\end{tabular}}

\newcommand{\metricheader}[9]{
\begin{table}[H]
\begin{tabularx}{\columnwidth}{lllXr}
\multicolumn{4}{l}{\textit{Name}: \textbf{#1}} &\\ \hline
\multicolumn{5}{l}{\textit{Synopsis}: \multicell{#2}} \\ \hline
\multicolumn{1}{l|}{\begin{tabular}[c]{@{}l@{}}\textit{Severity}:\\ #4\end{tabular}} & \multicolumn{1}{l|}{\begin{tabular}[c]{@{}l@{}}\textit{Effort}:\\ #5\end{tabular}} & \begin{tabular}[c]{@{}l@{}}\textit{Data Source}:\\ #6\end{tabular} \\ \hline
\end{tabularx}
\end{table}
\vspace{-0.8cm}
}

\newcommand{\HRule}[1][\medskipamount]{\par
  \vspace*{-0.3cm}
  \begin{adjustwidth}{-6pt}{0pt}
  \noindent\rule{\columnwidth}{0.2mm}\par
  \end{adjustwidth}
  \vspace*{-0.07cm}
}

\IEEEoverridecommandlockouts
\usepackage{tikz}
\newcommand\copyrighttext{%
  \footnotesize C. Matthies, T. Kowark, M. Uflacker and H. Plattner, ``\textit{Agile metrics for a university software engineering course},'' 2016 IEEE Frontiers in Education Conference (FIE), Erie, PA, USA, 2016, pp. 1-5. doi: \href{https://doi.org/10.1109/FIE.2016.7757684}{10.1109/FIE.2016.7757684} \\[4pt]
  Copyright \textcopyright 2016 IEEE. Personal use of this material is permitted.
  Permission from IEEE must be obtained for all other uses, in any current or future 
  media, including reprinting/republishing this material for advertising or promotional 
  purposes, creating new collective works, for resale or redistribution to servers or 
  lists, or reuse of any copyrighted component of this work in other works.
  }
\newcommand\copyrightnotice{%
    \begin{tikzpicture}[remember picture,overlay]
    \node[anchor=south,yshift=10pt] at (current page.south) {\fbox{\parbox{\dimexpr\textwidth-\fboxsep-\fboxrule\relax}{\copyrighttext}}};
    \end{tikzpicture}%
}

\begin{document}

\title{Agile Metrics for a University Software Engineering Course}

\author{\IEEEauthorblockN{Christoph Matthies, Thomas Kowark, Matthias Uflacker, and Hasso Plattner}
\IEEEauthorblockA{Hasso Plattner Institute, University of Potsdam\\
August-Bebel-Str. 88\\
Potsdam, Germany\\
Email: \{firstname.lastname\}@hpi.de}
}

\maketitle

% Render IEEE copyright notice
\copyrightnotice

\begin{abstract}
Teaching agile software development by pairing lectures with hands-on projects has become the norm. This approach poses the problem of grading and evaluating practical project work as well as process conformance during development.
Yet, few best practices exist for measuring the success of students in implementing agile practices. Most university courses rely on observations during the course or final oral exams. In this paper, we propose a set of metrics which give insights into the adherence to agile practices in teams. The metrics identify instances in development data, e.g. commits or user stories, where agile processes were not followed. The identified violations can serve as starting points for further investigation and team discussions.
With contextual knowledge of the violation, the executed process or the metric itself can be refined. The metrics reflect our experiences with running a software engineering course over the last five years. They measure aspects which students frequently have issues with and that diminish process adoption and student engagement. 
We present the proposed metrics, which were tested in the latest course installment, alongside tutoring, lectures, and oral exams.
\end{abstract}


\vspace{-0.1mm}

\section{Introduction}
In order to provide feedback to students and educators on how well Scrum and agile best practices are followed in a team, the day-to-day development process needs to be assessed.
We propose objective, automated \emph{conformance metrics} which can perform this assessment, complementing proven techniques, such as assessments by tutors or exams.
Conformance metrics rely on collected development data, e.g. commits or issues, which are created during regular development activities.
This ensures that established workflows do not need to be adapted and additional documentation overhead for students is avoided.
Conformance metrics measure to what degree an executed process is in agreement with previously defined ones, i.e. practices recommended by agile methodologies such as Scrum or XP.

Our approach detects and quantifies instances where the executed process deviates from the defined one. These instances, i.e. patterns in the collected data that do not comply with the details of the process, are referred to as \emph{violations}. Violations can reveal problem areas in the executed process, that need special attention.
This approach is comparable to test coverage tools and the Lint~\cite{Johnson1978} tool, which do not guarantee the absence of defects, but identify weaknesses.

\section{Conformance Metrics}
\begin{figure}[htb]
    \centering
    \includegraphics[width=0.9\columnwidth]{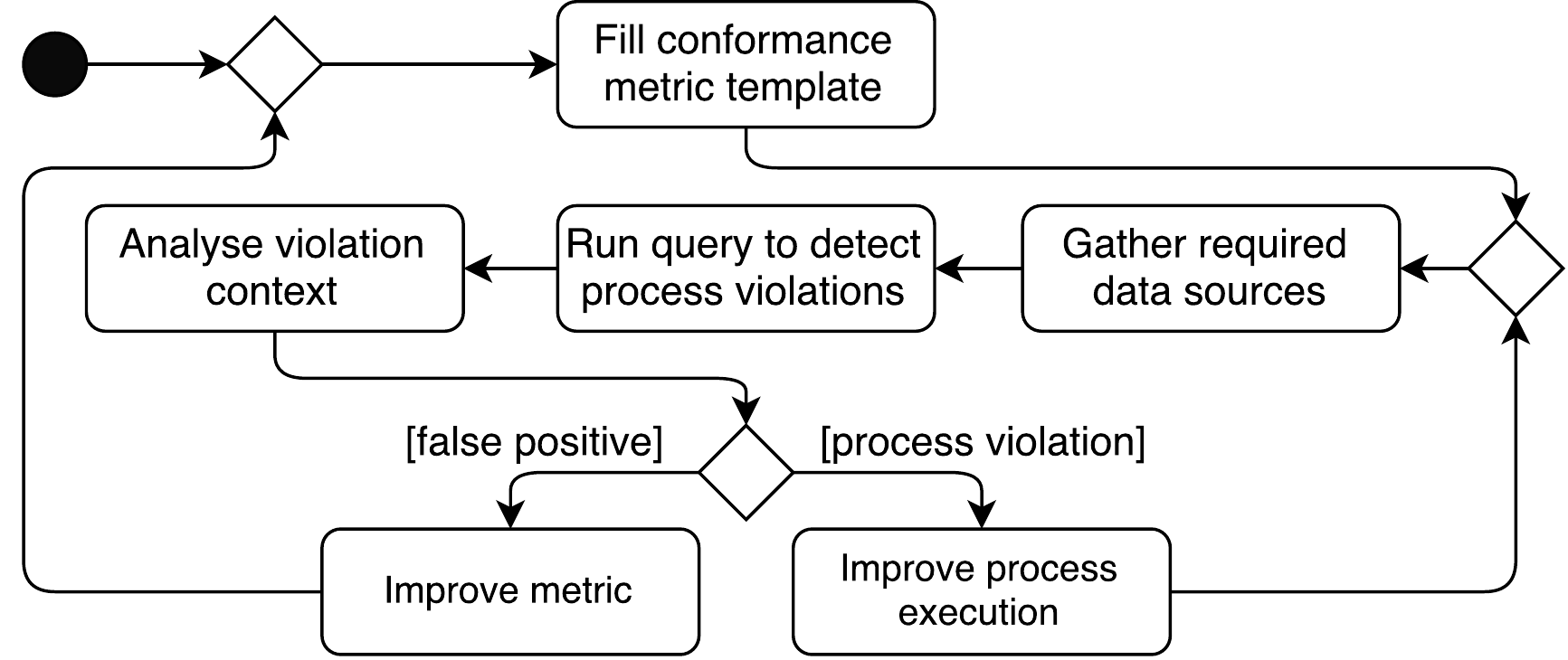}
    \setlength{\abovecaptionskip}{-4pt}
    \caption{Activity diagram  of the iterative lifecycle of conformance metrics.}
    \label{fig:process}
\end{figure}

Conformance metrics follow the iterative model described in Figure~\ref{fig:process}, adapted from Zazworka et al.~\cite{zazworka2010developers}: conformance metrics are defined, violations are detected, the context of these violations is researched and measures are taken to prevent future violations.

\subsection{Definition}
\label{sec:defconfmetrics}
In order to create a conformance metric there must be a common understanding of the practice that should be executed and measured. This involves both those who have knowledge and experience in agile development, i.e. the teaching staff as well a the team, who might have previous experience and personal preferences.
Agile methodologies such as XP or Scrum advise a multitude of practices, e.g. \enquote{no functionality is added early} or \enquote{all code must have unit tests}. Sletholt et al.~\cite{Sletholt2012} mention 35 main ones, which can serve as a starting point to select practices that are applicable in the context of a certain project.
If a process is considered relevant enough to be measured and a common understanding of its details is found, this knowledge should be recorded in the form of the process conformance template, based on previous work~\cite{zazworka2010developers}, see Table~\ref{table:template}.

\begin{table}[htb]
\centering
\begin{tabularx}{\columnwidth}{lX}
\textit{Name} & The unique, descriptive identifier of the metric. \\ \hline
\textit{Synopsis} & A short description of the type of violations the metric measures, e.g. \enquote{commits without tests}. \\ \hline
\textit{\multicell{Descrip-\\tion}} & Overview of the expected process, i.e. the practice which should be followed, and its advantages, with references to literature. A description of what constitutes a violation of this process should be included. \\ \hline
\textit{\multicell{Data\\sources}} & A list of data sources the metric requires and which the \emph{query} is based on, e.g. code repositories or issue trackers. \\ \hline
\textit{Query} & Steps needed to extract violations from the \emph{data sources}. Ideally, these steps can be automated, e.g. as a database query. \\ \hline
\textit{\multicell{Rating\\ function}} & Function that maps detected violations into a numerical score, indicating the degree of mismatch between the executed process and the one detailed in the \emph{description}. \\ \hline
\textit{Pitfalls} & Description of what the metric does \emph{not} measure, e.g. limitations or possible misconceptions about the results of the metric.\\ \hline
\textit{\multicell{Cate-\\gories}} & Topics in the domain of agile software development the metric attempts at measuring, e.g. \enquote{XP practices} \\ \hline
\textit{Effort} & How much effort collecting violations and calculating a score requires. Either low, medium or high, e.g. using an automated process on existing data sources is \enquote{low} effort. Low effort metrics should be implemented first.\\ \hline
\textit{Severity} & Importance in the context of the project's agile development process. How severe violations found by this metric are. Either informational, very low, low, normal or high. \\ \hline
\end{tabularx}
\caption{Conformance metric template.}
\label{table:template}
\end{table}

\paragraph{Description}
The only requirement of a metric's description is that it is detailed enough to allow defining what patterns in the collected data constitute a violation and which do not. This means the description may, but is not required to, follow formal definitions.

\begin{figure}[!ht]
    \centering
    \includegraphics[width=\columnwidth]{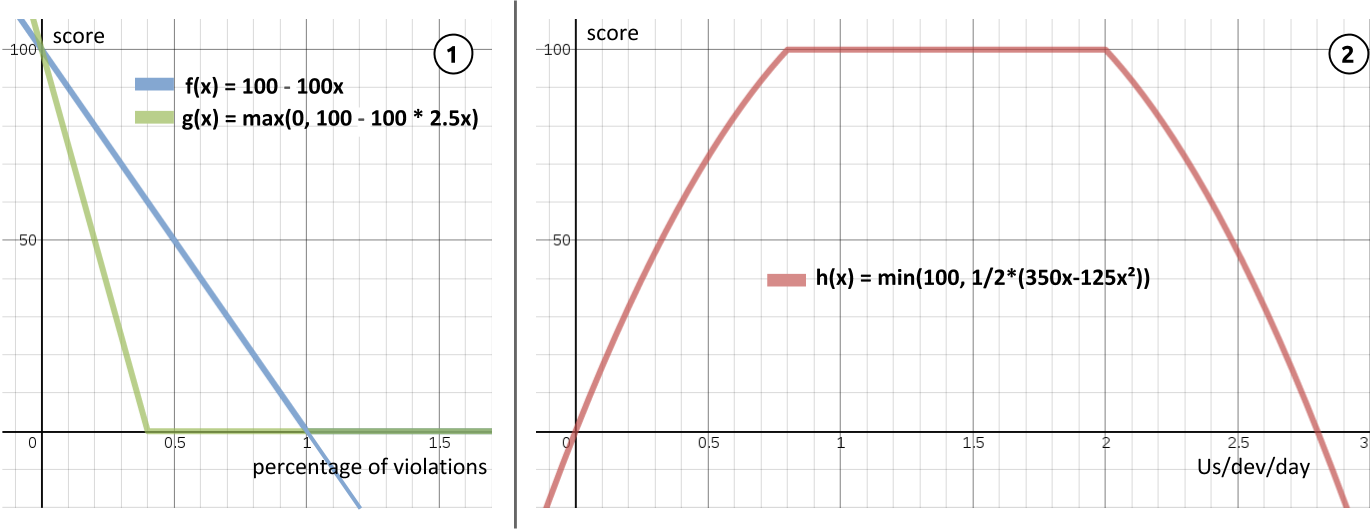}
    \setlength{\abovecaptionskip}{-4pt}
    \caption{Graphs of examples of the two main types of rating functions, threshold (1) and cut-off parabola (2), mapping the output of conformance metrics to a score between 0 and 100.}
    \label{fig:graphs}
\end{figure}

\paragraph{Score Calculation}
In order to allow users a quick overview, the results of a metric are summarized using a rating function.
It maps the violation details returned by the query into the more abstract form of a score, bounded by a high and low value
We employed percentages from 0 to 100, where 100 indicates that no violations were detected and 0 that the described practice was not followed at all.
Numerical values allow the development of a metric over time to be visualized, as well the results of individual metrics to be summed into an overall team score in an iteration, e.g. a sprint.
Two main types of rating function were used:
\begin{enumerate}
  \item \emph{Threshold function}, a linearly decreasing function, returning 0 for inputs larger than a threshold, e.g. $g(x)$ in Figure~\ref{fig:graphs}.1.
  \item \emph{Cut-off parabola}, returning the perfect scores for a range of optimal input values, while results for values outside of the optimal range quickly fall, e.g. $h(x)$ in Figure~\ref{fig:graphs}.2.
\end{enumerate}

\paragraph{Classification}
The severity of a conformance metric reflects the relative importance of the metric for the process. The severity determines the weight with which the metric influences the overall score of a team. We identified three main factors, which influenced the initial severity rating:
\begin{enumerate}
  \item \textbf{Importance in agile methodologies} Core elements of agile methodologies, e.g. working in iterations, were considered to be of high severity.
  Practices that are found in multiple methodologies, e.g. Scrum and XP, were rated more severe.
  \item \textbf{Importance for team} Agile development practices need to be adapted to a team's context. Even established practices can be of low importance to a particular team.
  \item \textbf{Confidence in generally established values} While there is some level of consensus on agile core practices~\cite{Sletholt2012}, there is often little consensus on what values constitute a violation.
  Until reasonable values for a team are found through iterative refinement, the severity of metrics would initially be low.
\end{enumerate}

\subsection{Query Execution}
After development data has been collected from the defined data sources, the query part of the metrics can be used to extract process violations, see Listing~\ref{lst:query}.
The rating function then outputs a list of process violations by teams and iteration as well as a score, indicating the seriousness of the extracted violations.

\begin{listing}[ht]
    \begin{minted}[fontsize=\scriptsize, frame=single]{sql}
MATCH (m:Milestone)-[]-(i:Issue)-[]-(l:Label)
WHERE m.title = "Sprint 12"
and m.due_on<timestamp() AND i.current_state="open"
WITH m, m.title as Sprint, collect(DISTINCT i) AS Issues,
    count(DISTINCT i) AS Amount
MATCH m-[:milestone]-(j:Issue)
WITH Issues, count(j) AS Total, Amount, m.title as Sprint
RETURN Sprint, Amount, Issues, Total,
    Amount/Total AS Percent
    \end{minted}
    
\begin{tabularx}{\columnwidth}{|l|l|l|l|X|}
\hline
\textbf{Sprint} & \textbf{Amount} & \textbf{Issues} & \textbf{Total} & \textbf{Percent} \\ \hline
Sprint 12 & 2 & \#129, \#135 & 10 & 0.2 \\ \hline
\end{tabularx}
    %\vspace*{-0.5cm}
    \caption{Example of a Neo4J graph database query~\cite{holzschuher2013performance}, as well as an example result, of user stories that were part of sprints before Sprint 12 but are not marked as done.}
    \label{lst:query}
\end{listing}

\subsection{Context Analysis}
\label{sec:research}
In order to gather information about the quality of violations, e.g. identifying false positives, additional information on detected violations is needed.
For example, the change history of a user story can be viewed or directly discussed with the involved developers.

\subsection{Improvement}
\label{sec:improve}
A major part of the lifecycle of conformance metrics is their continuous improvement; this means ensuring that the metrics fit the project and team context.
Especially the amount of false positives, i.e. process violations that were detected but do not actually pose a problem for a development team, need to be minimized.
The amount of real violations, i.e. true positives, can be reduced by modifying the query or process description, thereby adjusting the metric to better fit the executed process, or by making sure the defined process is followed more closely, e.g. by additional training or tutoring~\cite{Matthies2016a}.

\noindent
The knowledge that was gathered in the improvement step can then be made explicit again in the first step of the lifecycle, by editing the conformance template.

\section{Conformance Metrics Details}
While the conformance metrics presented in this section are based on common agile practices, they are often tailored to the specific set of Scrum, agile and organizational practices that we focused on in student courses.
We therefore do not include the query or the rating function, as these are not universal. They are available in a open-source licensed repository\footnote{\url{https://github.com/hpi-epic/ScrumLint}}.
The metrics in this section are divided by their categories.

\paragraph{XP Practices}
Metrics measuring violations of Extreme Programming development practices.

\metricheader{Collective Code Ownership}
% Synopsis
{Code which is edited heavily by few developers.}
% Category, Severity, Effort, Data Source
{XP Practices}{Normal}{Low}{Version control}
%Rating function
{$max(0, (100 - \#violations * weight))$}
% comment, id
{$\#violations$ = amount of files with $threshold_{e}$ edits\\by $threshold_{a}$ authors.}{1}
\noindent
\begin{adjustwidth}{6pt}{0pt}
\vspace{4pt}
\textit{Description:} Collective Code Ownership, as defined by Beck~\cite{Beck1999} states that \enquote{every programmer improves any code anywhere in the system at any time if they see the opportunity}. It is one of the core extreme programming practices. Closely related is the \enquote{bus number}~\cite{ricca2011difficulty}, which is the number of developers that a project would need to lose to halt its progress. It measures the concentration of knowledge about software components in individual team members. Following the practice of Collective Code Ownership can help every developer work on any user story.
This metric finds files that had many edits by only few authors. The more of these there are, the less the practice of Collective Code Ownership was followed. 
\end{adjustwidth}

\metricheader{Test-Later Development}
{Increasing code complexity while decreasing\\code coverage.}
{XP Practices}{Normal}{Medium}{version control, code coverage stats}
{$max(0, 100-(\#violations\div\#commits*100*weight))$}
{$\#violations$ = amount commits increasing complexity \& decreasing coverage.}{2}
\noindent
\begin{adjustwidth}{6pt}{0pt}
\vspace{4pt}
\textit{Description:} In TDD, an automated test is written before the code that makes it pass. This is followed by a refactoring step. Following TDD can have a positive effect on system design and assures that all code is always tested~\cite{Madeyski10}. Kniberg states 
%This [TDD], to me, is more important than both Scrum and XP. 
\enquote{You can take my house and my TV and my dog, but don’t try to stop me from doing TDD!}~\cite{Kniberg2007}. TDD is also related to the XP practice of \emph{YAGNI} (you ain't gonna need it)~\cite{Jeffries2001}. Tests act as a reminder to work on the current story.
This metric identifies commits where TDD was not followed, i.e. commits which introduced additional complexity to the system, but led to decreased code coverage. \newline
\end{adjustwidth}

\metricheader{Huge User Stories}
{User stories that are unusually large.}
{XP Practices}{Low}{Low}{User story tracker}
{$max(0, 100-(\#violations*weight))$}
{$\#violations$ = amount user stories larger than \\$threshold_{length}$ multiplied with average length of user stories or with \\$threshold_{check}$ multiplied with average amount of checkboxes of user stories.}{3}
\begin{adjustwidth}{6pt}{0pt}
\vspace{4pt}
\textit{Description:} User stories should be small enough to get a quick overview of the work to be done, but should contain enough information to allow developers to estimate it. The text of a user story should fit on an index card~\cite{Cohn2004}.
If a user story is much longer than the average this might be an indicator that it is too large, was hard to estimate and should be split~\cite{Jeffries2001}.
The user stories identified by this metric are significantly above the average length of stories in the sprint or have many times the amount of tasks of other stories. \newline
\end{adjustwidth}

\paragraph{Backlog Maintenance}
Metrics attempting to measure violations concerning both product and sprint backlogs.

\metricheader{One Story, Multiple Backlogs}
{User stories that were in multiple sprint backlogs.}
{Backlog Maintenance}{High}{Low}{User story tracker}
{$max(0, 100-(\frac{\#violations}{\#totalUS}*100*AvgInSprints*weight))$}
{$\#violations$ = amount of user stories in more than  $threshold_{amount}$ sprints,\\$\#totalUS$ = total amount of user stories in the Sprint Backlog,\\ $AvgInSprints$ = average amount of sprint backlogs the violations were in.}{4}
\begin{adjustwidth}{6pt}{0pt}
\vspace{4pt}
\textit{Description:} Ideally, a sprint backlog contains exactly as many user stories as the team can complete in an iteration~\cite{Schwaber2011}, so that at the end of the sprint all user stories in the sprint backlog are completed. This ensures the ability to plan the software's development and enables teams to build on the finished functionality in the next sprint.
However, sometimes, at the end of the sprint not all stories conform to the \enquote{Definition of Done}~\cite{Kniberg2007}. These user stories are then carried over to the next sprint, if the product owner still considers them a priority. A story that spans multiple sprints can be a blocker for other teams that depend on it.
This metric identifies user stories that were assigned to the sprint backlog of multiple sprints. The percentage of offending \enquote{neverending} stories should be minimal.
\newline
\end{adjustwidth}

\metricheader{Duplicates}
{User Stories which are suspected duplicates.}
{Backlog Maintenance}{Very Low}{Low}{User story tracker}
{$max(0, 100-((\#duplicates\div\#totalUS)*100*weight))$}
{$\#duplicates$ = amount of user stories in the tagged as duplicates.\\ $\#totalUS$ = total amount of user stories.}{5}
\begin{adjustwidth}{6pt}{0pt}
\vspace{4pt}
\textit{Pitfalls:} This metric relies on developers or teaching staff tagging user stories as duplicates. There might be additional duplicates that were not tagged.
\HRule
\textit{Description:} User stories are the main tool of specifying what will be done in a sprint~\cite{Cohn2004}. User stories should not overlap in described functionality, as there is the risk of features being developed twice if these user stories are given to different teams in the same sprint.
This metric identifies user stories that were marked as possible duplicates by developers. \newline
\end{adjustwidth}

\paragraph{Developer Productivity}
Metrics dealing with topics such as how work is structured, how it is assigned and the workload of developers.

\metricheader{At the Last Minute}
{Commits shortly before sprint deadline.}
{Dev. Productivity}{Normal}{Low}{Version control}
{$max(0, (100 - (\dfrac{\#violations}{\#totalCommits}*100*weight) )$}
{$\#violations$ = amount of commits made within $x$ minutes to sprint end.\\ $\#totalCommits$ = amount of total commits.}{6}
\begin{adjustwidth}{6pt}{0pt}
\vspace{4pt}
\textit{Description:} Work in an agile project should follow a \enquote{sustainable, measurable, predictable pace}~\cite{wells99} and overtime should be avoided~\cite{Beck1999}. The software at the end of the sprint should be as completed, tested and integrated as possible. If work is slanted towards the end of the sprint and code is committed at the very last minute, this can cause a range of problems: Scrum meetings might be ineffective, due to lack of content, blockers for or by other teams can not be communicated in a timely fashion and code review through other developers becomes more difficult.
This metric measures commits that were made during the last minutes before sprint deadline. The more of these there are, the less likely it is that a sustainable pace was followed. \newline
\end{adjustwidth}

\metricheader{No Committing}
{Average amount of commits per developer.}
{Dev. Productivity}{Normal}{Low}{Version control}
{$min(100, \#commits\div\#developers*weight)$}
{$\#commits$ = amount of commits,\\ $\#developers$ = amount of developers in a team.}{7}
\begin{adjustwidth}{6pt}{0pt}
\vspace{4pt}
\textit{Description:}  The rule of \emph{Check in Early, Check in Often} is encourages small patch sizes~\cite{Bosu14}. Jeff Atwood, co-founder of Stack Overflow, considers it a \enquote{golden rule of source control}. He states that from a team member's viewpoint, \enquote{if the code isn't checked into source control, it doesn't exist}~\cite{atwood08}. Committing finished functionality frequently is also a requirement for continuous integration and delivery called for in the principles of the agile manifesto~\cite{beck2001agile}. Committing often allows coworkers to build on functionality, review the code and makes version control and merging easier. 
This metric measures the average amount of commits that were made by the developers of a team over the course of a sprint. Generally, the more commits were made, the better, however, they should represent working increments of the software. \newline
\end{adjustwidth}

\metricheader{Daily User Story Amount}
{Average amount of user stories a developer is\\assigned per day.}
{Dev. Productivity}{Low}{Low}{User story tracker}
{\\$min(100, weight_{a}*quota-weight_{b}*quota^{2})$}
{$quota = \#devs\div\#sprintBacklog\div sprintLength$,\\ where $\#devs$ = amount of developers,\\ $\#sprintBacklog$ = size of the Sprint Backlog, \\ $sprintLength$ = length of the sprint in days.}{8}
\begin{adjustwidth}{6pt}{0pt}
\vspace{4pt}
\textit{Description:} User stories should conform to the \emph{INVEST} acronym (independent, negotiable, valuable, estimable, small, testable). Small has been defined to  mean a few person-days to a few person-weeks~\cite{wake2003invest}. Cohn does not state absolute values but explains that \enquote{the ultimate determination of whether a story is appropriately sized is based on the team, its capabilities, and the technologies in use}~\cite{Cohn2004}. While the amount of user stories a developer should be able to finish per day is hard to state generally, working on multiple stories per day results in increased context switching overhead~\cite{johnson2003beyond}.
This metric measures the average amount of user stories a developer would have to finish every day, given constant productivity. If this number is high, it is possible that the requirements of the Definition of Done, deployment, communication and context switching overhead were underrated and the Sprint Backlog is too full. \newline
\end{adjustwidth}

\metricheader{Fast pull requests}
{Pull requests that were closed quickly\\without comments.}
{Dev. Productivity}{High}{Low}{Pull Requests}
{\\$max(0, 100-(\#violations\div\#totalPRs*100))$}
{$\#violations$ = amount of pull requests that were closed\\quickly, $\#totalPRs$ = total amount of pull requests.}{9}
\begin{adjustwidth}{6pt}{0pt}
\vspace{4pt}
\textit{Description:} Pull requests can be a tool to help inform team members what functionality is added in a collection of commits. It allows team members and stakeholders to comment and perform code review. According to Boehm and Basili, code reviews by peers catch around 60\% of defects~\cite{Boehm2001}. Furthermore, continuous integration services can run the proposed changes, making sure all tests pass, before code is merged. If pull requests are merged in a short timespan without anyone commenting, this hints at many of these possibilities remaining unused.
This metric identifies pull requests that were closed quickly and had no comments. The more of these \enquote{speedy pulls} are found, as a percentage of all pull requests, the worse the score. \newline
\end{adjustwidth}

\section{Conclusion}
Using the presented conformance metrics, violations for all teams in all sprints could be extracted from data gathered in the 2014/15 installment of our undergraduate agile software development course with 38 participants.
As the presented metrics rely solely on data, they allow a more unbiased view of teams.
By providing the offending development artifacts, the root causes of violations could be established, such as bad communication between teams involved with user management.
All metrics could be used to identify instances where tutor intervention would have been helpful.
In some cases, severe violations were found that were missed by tutors, such as a very complex, wrongly prioritized user story, that had been in the sprint backlog of all sprints.
As it was one user story among hundreds it was missed by manual analyses.
As such, we see this data-driven approach as a good supplement to the usually employed assessment techniques in undergraduate student software engineering capstone projects.
Future work includes employing the presented metrics in the next installments of project courses as well as continuously improving them in order to better reflect project reality and eliminate false positives.

\newpage

%\bibliographystyle{IEEEtran}
\bibliographystyle{IEEEtran}
\bibliography{IEEEabrv,library}

\end{document}